\documentclass[12pt,amsmath,amssymb,pra]{revtex4}

\usepackage{graphicx}
\usepackage{amsmath}
\usepackage{amsfonts}%
\usepackage{amssymb}%

\setcitestyle{numbers,square}
\begin{document}

\title{Chapter 4: Analog quantum simulation}

\maketitle

\section{Ultracold atoms}

We have seen in the previous chapters that classical simulation
methods do not work for many interesting many-body problems, most
notably involving fermions or frustrated interactions in two or three
spatial dimensions. The root of this difficulty arises from the
Hilbert space dimension growing exponentially with the size of the
system. Therefore, it looks like we need to use a simulator in which
the available resources scale in a similar fashion, meaning the
simulator has to be a quantum system itself! This was first recognized
by Richard Feynman in 1982 \cite{Feynman1982}
and is widely thought to mark the birth of quantum simulation. A good
quantum simulator is a physical system which can be well controlled so
it can mimic basically any other quantum system that might be of
interest. Here, ultracold atoms have been proven to be very
attractive: 
\begin{enumerate}
\item  Individual atoms are well isolated from the environment.
\item  Their properties can be widely tuned with electric fields, magnetic
fields, microwaves, and lasers.  
\item  Theoretical predictions can be
derived from first principles.  
\item Ultracold temperatures in the nanokelvin range are accessible
  using laser cooling methods.
\end{enumerate}

Ultracold atoms can be either bosonic or fermionic, which depends on the total spin of all constituents. As neutral atoms contain as many protons as electrons, their contribution to the spin will always be an integer number. Consequently, the statistical properties are set by the number of neutrons, i.e., we have bosons for an even number of neutrons and fermions for an odd number of neutrons. A prominent example of a bosonic atom is $^{87}\textrm{Rb}$ (50 neutrons), while an important fermionic isotope is $^{40}\textrm{K}$ (21 neutrons). Note that in the case of potassium, we can actually have both a fermionic and a bosonic ($^{39}\textrm{K}$) isotope.

A many-body system of ultracold atoms is most coveniently described in the formalism of the second quantization. Let us first consider spinless bosons, for which we choose a Fock space for each possible mode as a suitable basis, i.e., eigenstates of the particle number operator for each mode, $\{|n\rangle_k\}$. The total Hilbert space is then formed as the tensor product of all these single-mode Fock spaces, $\mathcal{H}=\otimes_k \mathcal{H}_k$. The creation and annihilation operators for the Fock states satisfy the commutation relations known from the harmonic oscillator:
\begin{eqnarray}[a_k,a_l]&=&[a_k^\dagger,a_l^\dagger] = 0\\{} [a_k,a^\dagger_l]&=&\delta_{kl}.\end{eqnarray}Additionally, the annihilation operator vanishes when applied to the vacuum state of the mode, $a_k|0\rangle_k = 0$, and the number operator for each mode is simply $n=a_k^\dagger a_k$.

For fermions, the commutation relations are replaced by anticommutation relations:
\begin{eqnarray}\{a_k,a_l\}&=&\{a_k^\dagger,a_l^\dagger\} = 0\\\{a_k,a^\dagger_l\}&=&\delta_{kl}.\end{eqnarray}A consequence of these relations is the Pauli exclusion principle, which states that two fermions cannot occupy the same state, i.e., $a_k^\dagger a_k^\dagger |0\rangle_k = 0$. As operators between different modes no longer commute (they anti-commute instead), we have to define an ordering of the modes. Usually, a multi-mode Fock state is defined as
\begin{equation}|n_1,n_2,\ldots n_i,\ldots\rangle = (a_i^\dagger)^n_i (a_2^\dagger)^n_2 (a_1^\dagger)^n_1 |0,0,\ldots,0,\ldots\rangle,\end{equation}i.e., the first mode appears on the right. To understand the subtle point about the importance of ordering, consider this two-mode example:
\begin{equation}a_1^\dagger |0,1\rangle = a_1^\dagger a_2^\dagger|0,0\rangle = - a_2^\dagger a_1^\dagger|0,0\rangle = -|1,1\rangle.\end{equation}
Accounting for spin is straightforward in this formalism and only requires to replace the operators $a_k$ (and its hermitian conjugate) by $a_{k,\sigma}$, where $\sigma$ is the spin variable. We can also define field operators, which will allow us to express observables and the Hamiltonian in a convenient way:
\begin{eqnarray}\psi(r)&=& \frac{1}{\sqrt{V}}\sum_k e^{ikr} a_k\\\psi(r)^\dagger &=& \frac{1}{\sqrt{V}}\sum_k e^{-ikr} a_k^\dagger,\end{eqnarray}where $V$ is the quantization volume. Depending on the statistics of the creation and annihilation operators, the field operators will obey bosonic or fermionic statistics as well. Since the field creation operator is the Fourier transform of the mode creation operator $a_k^\dagger$, $\psi^\dagger(r)$ acting on the vacuum will simply create a particle at position $r$. We can use this to define the density operator
\begin{equation}n(r) = \psi^\dagger(r)\psi(r),\end{equation}which by integration over the full space gives the total particle number, as expected:
\begin{equation}N = \int\textup{d}r \psi^\dagger(r)\psi(r) = \int\textup{d}r \frac{1}{V}\sum_{k,k'} e^{i(k-k')r} a^\dagger_k a_k = \sum_k a^\dagger_k a_k.\end{equation}
We are now in the position to formulate the Hamiltonian for the many-body system of ultracold atoms. The kinetic energy is diagonal when using creation and annihilation operators,
\begin{equation}H_\mathrm{kin} = \sum_k \frac{\hbar^2k^2}{2m} a_k^\dagger a_k = \int \textup{d}r \frac{\hbar^2}{2m} [\nabla \psi^\dagger(r)][\nabla \psi(r)].\end{equation}Likewise, the energy associated to an external potential (e.g., describing a trapping potential) is given by
\begin{equation}H_\mathrm{pot} = \int \textup{d}r n(r) V(r) = \int \textup{d}r \psi^\dagger(r)\psi(r)V(r).\end{equation}
Finally, the only thing that is missing is the interaction term. In most cases the interaction will only depend on the spatial separation of two particles and hence can be expressed as
\begin{equation}H_\mathrm{int} = \frac{1}{2}\int \textup{d}r\textup{d}r' \psi^\dagger(r)\psi^\dagger(r')V_\mathrm{int}(r-r')\psi(r')\psi(r) = \frac{1}{V} \sum_{k.k',q} \tilde{V}_q a^\dagger_{k+q} a^\dagger_{k'-q} a_{k'} a_k,\end{equation}where $\tilde{V}_q$ is the Fourier transform of the interaction potential. Note that for fermions, the ordering of the creation and annihilation operators is important, while for bosons it only matters that creation operators appear to the left of annihilation operators.

To develop an understanding of the interactions arising between ultracold atoms, we first have to look into their electronic structure in some detail. We focus on alkali atoms like Rb as they are the most commonly used ones in ultracold atoms experiments. In their electronic ground states, two Rb atoms have their single valence electron in the $|5s\rangle$ state, i.e., $|5s,5s\rangle$ in the two-atom basis. The atoms are neutral and their magnetic dipole moments of $1\,\mu_B$ are negligible, so the dominant contribution comes from a van der Waals interaction behaving as $-C_6/r^6$, which is a second-order off-resonant dipole-dipole interaction. Here, the dominant term comes from the dipole transitions to the first electronic excited state, resulting in
\begin{equation}C_6 \sim \frac{\langle 5_s,5_s|x_1 x_2|5_p,5p\rangle\langle 5p,5p |x_1 x_2|5s,5s\rangle}{2\Delta},\end{equation}where  $x_1$ and $x_2$ are the coordinates for the valence electrons and $\Delta$ is the energy gap to the first excited state. Using the known values for Rb \cite{Steck2001},  we obtain a value for the van der Waals coefficient of $C_6 = 3100 \mathrm{a.u.}$, which is not too far off from the experimentally measure value of $C_6 = 4707 \mathrm{a.u.}$ \cite{Claussen2003}. This large value of the van der Waals coefficient can be understood as the single valence electron in alkali atoms behaving almost hydrogenic (in hydrogen, the excitation gap $\Delta$ vanishes for $ns-np$ transitions).

While the van der Waals interaction gives a correct description at large separation, at short ranges core repulsion will kick in, leading to a molecular potential supporting many bound states. However, in many cases the short range details are not important. We consider the two-atom Schrödinger equation,
\begin{equation}\left[-\frac{1}{2\mu}\Delta + V(\textbf{r})\right]\psi(\textbf{r}) = E\psi(\textbf{r}),\end{equation}with $\mu$ being the reduced mass. At long distances, we can look at its asymptotic behavior \cite{Friedrich2006},
\begin{equation}\psi(\textbf{r}) = e^{ikx} + f(\theta,\phi)\frac{e^{ikr}}{r}.\end{equation}This first term in this expression describes an incoming plane wave, while the second term describes an outgoing spherical wave after the scattering event between the two atoms. For radially symmetric interaction potentials, the dependence on the azimuthal angle drops out and the scattering amplitude $f(\theta)$ can be expressed with the help of Legendre polynomials as \cite{Friedrich2006}
\begin{equation}f(\theta) = \sum_{l=0}^\infty f_l P_l(\cos \theta).\end{equation}In this form, we have an expression in terms of different partial waves with angular momentum $l$. In analogy to atomic angular momentum, the $l=0$ channel is called $s$--wave scattering, the $l=1$ channel is called $p$-wave scattering and so on. The partial wave amplitudes $f_l$ take the form \cite{Friedrich2006}
\begin{equation}f_l = \frac{2l+1}{2ik}(e^{2i\delta_l}-1),\end{equation}where $\delta_l$ denotes the scattering phase shift. At low temperatures, we are interested in how the scattering amplitude behaves in the limit $k\to 0$. In particular, for the van der Waals interaction we have
\begin{equation}\delta_0 = -k a_s\end{equation}\begin{equation}\delta_1 \sim k^3\end{equation}\begin{equation}\delta_l \sim k^4 \;\;\; l>1\end{equation}Here, we have introduced the $s$-wave scattering length $a_s$, in which the details of the scattering process have been absorbed. The typical scale is given by the effective range $r_e$,
\begin{equation}a_s \sim r_e =  (2\mu C_6)^{1/4},\end{equation}however, close to resonances in the molecular channel (so-called Feshbach resonances), the scattering length will substantially deviate from this typical scale \cite{Bloch2008}. Consequently, at low $k$, we can expect $s$-wave scattering to be the most relevant term. Then, we can use the analytic expression for the scattering amplitude
\begin{equation}f(k) = \frac{1}{-1/a_s + r_e k^2/2 - ik}.\end{equation}As long as the energy scales are lower than the characteristic scale $1/(2\mu r_e)^2$, we can neglect the contribution from the effective range and are in the regime where the scattering is completely described by the scattering length (typical temperatures: $< 1\,\mathrm{mK}$). This is the regime of ultracold atoms.

We are now looking for an interaction potential in position space that reproduces this scattering behavior. This is achieved by the pseudopotential
\begin{equation}V(\textbf{r}) = \frac{4\pi a_s}{2\mu} \delta(\textbf{r})\frac{\partial}{\partial r} r.\end{equation}However, the term involving the partial derivative is only important when the wave function is singular at the origin and can be neglected in most cases. Then, the interaction term in the many-body Hamiltonian simplifies to 
\begin{equation}H_\mathrm{int} = \frac{1}{2}\int \mathrm{d}r\mathrm{d}r' \psi^\dagger(r)\psi^\dagger(r')V_\mathrm{int}(r-r')\psi(r')\psi(r) = \frac{g}{2}\int \mathrm{d}r \psi^\dagger(r)\psi^\dagger(r)\psi(r)\psi(r).\end{equation}Note that for fermions this term vanishes due to the Pauli exclusion principle. In this case, the dominant scattering occurs in the much weaker $p$-wave channel.

\section{Optical lattices}

The ground state of the bosonic many-body Hamiltonian with a contact interaction is a Bose-Einstein condensate, i.e., even in the presence of strong interactions, the $k=0$ mode gets macroscopically occupied, i.e., 
\begin{equation}\lim_{N\to\infty}\langle a^\dagger_0 a_0\rangle/N = n_c > 0,\end{equation}where $n_c$ is the condensate fraction. Closely related is the concept of off-diagonal long-range order (ODLRO),
\begin{equation}\lim_{r\to\infty}\langle \psi^\dagger(r)\psi(0)\rangle = n_c,\end{equation}from which we see that a Bose-Einstein condensate is characterized by phase coherence.

If we want to use an ultracold Bose gas as a quantum simulator for typical many-body problems arising in condensed matter physics, we should put the atoms into a periodic potentials similarly how it is done for electrons in a solid state lattice provided by the atomic nuclei. Such periodic structures can be efficiently realized for ultracold atoms using laser potentials. In particular, we consider near-resonant light acting on Rb atoms in their electronic ground state. Then, the laser will create a perturbation of the form
\begin{equation}V = d E_0 \cos(\omega t-k_Lx) + \mathrm{h.c.}.,\end{equation}
where $d is$ the dipole operator, $E_0$ is the strength of the electric field created by the laser, $\omega$ is the laser frequency, and $k_L$ its wavevector. As previously, we assume that only a single transition to one of the $5p$ is relevant. Then, the potential can be cast onto a two-level problem of the form
\begin{equation}V = \Delta|5p\rangle\langle 5p| + [\Omega \cos(\omega t - k_Lx)|5s\rangle\langle 5p| + \mathrm{h.c.}],\end{equation}where we have introduced the Rabi frequency $\Omega = d_{5s-5p} E_0$. We now go into the rotating frame of the laser field, i.e., we make the transformation $|5p\rangle \to |5_p\rangle \exp(i\omega t)$. Inserting this into the Schrödinger equation shifts the $5p$ level by the frequency $\omega$ and leads to the detuning $\delta = \Delta-\omega$. In the rotating-wave approximation, we neglect fast rotating terms on the order of $2\omega$ and obtain the effective Hamiltonian
\begin{equation}H = \delta |5p\rangle \langle 5p| + \left[\frac{\Omega}{2}\cos(k_L x)|5s\rangle\langle 5p| + \mathrm{h.c.}\right].\end{equation}If the detuning is much larger than the Rabi frequency, we can integrate out the excited state in second-order perturbation theory and obtain the effective potential for the ground state atoms,
\begin{equation}V(r) = \frac{\Omega^2}{2\delta}\cos^2(k_L x).\end{equation}

We are now interested in the dynamics of ultracold atoms in such an optical lattice potential. We first look at the behavior of a single atom. The laser potential is periodic, hence the solution will be given by Bloch waves of the form
\begin{equation}\phi_{nk}(r) = e^{ikr}u_{nk}(r),\end{equation}where $u_{nk}(r)$ are periodic functions. The Bloch wavefunctions are characterized by a quasi-momentum $k$, restricted to the first Brillouin zone of the reciprocal lattice, and a band index $n$. Dealing with Bloch waves is often inconvenient, however, therefore it is usually better to work with Wannier functions $w_n(r-R_i)$, which are well localized in space around a particular lattice minimum $R_i$. They are given by the Fourier transform of the Bloch waves over the first Brillouin zone,
\begin{equation}w_n(r-R_i) = \int\frac{\mathrm{d}k}{v}\phi_{nk}(r),\end{equation}where $v$ is the volume of the first Brillouin zone. Note that the definition of the Wannier functions is not unique, as the Bloch waves are defined only up to an arbitrary phase. We can, however, require the orthonormality relation,
\begin{equation}\int \mathrm{d}r w_n^*(r-R_i)w_{m}(r-R_j) = \delta_{mn}\delta_{ij}.\end{equation}We can also express the field operators in terms of the Wannier functions,
\begin{equation}\psi(r) = \sum_{R_i,n} w_n(r-R_i)a_{R,n}.\end{equation}Here, the operator $a_{R,n}$ annihilates are particle in a Wannier state corresponding to the lattice site $R_i$ and the Bloch band $n$. For reasonably deep lattices, the bands are well separated, and we may restrict our analysis to the lowest Bloch band.

Assuming separable lattice potentials, the problem reduces to the one-dimensional Schrödinger equation
\begin{equation}\left(-\frac{1}{2m}\frac{\mathrm{d}^2}{\mathrm{d}x^2} - \frac{V_0}{2} \cos 2k_L x \right)\psi(x) = E\psi(x).\end{equation}Using the substitutions $v=k_Lx$ and $q = V_0/(4 E_r)$, where $E_r = k_L^2/2m$ is the recoil energy of the laser, this can be mapped onto a Mathieu equation,
\begin{equation}\frac{\mathrm{d}^2}{\mathrm{d}x^2} y(v) + (E-2q\cos 2x) y(v)=0.\end{equation}The Mathieu equation has symmetric solutions with eigenvalues $a_r$ and antisymmetric solutions with eigenvalues $b_r$. Here, the lowest eigenvalue in the lowest band corresponds to the eigenvalue $a_0$, while $b_1$ is the eigenvalue of the highest state in the first band. Then, in the limit of deep optical lattices ($q \gg 1$) we can use an (unproven) relation to obtain the width $W$ of the first Bloch band \cite{Abramowitz1972},
\begin{equation}W = b_1-a_0 = \frac{16}{\sqrt{\pi}} \left(\frac{V_0}{E_r}\right)^{3/4} e^{-4\sqrt{V/E_r}}.\end{equation}In the lowest band, the dispersion relation is given by
\begin{equation}E_k = -2J\cos ka\end{equation}where $a=\pi/k_L$ is the lattice spacing and the characteristic energy scale $J$ is related to the bandwidth by $W=4J$. We can also express the Hamiltonian using the Wannier basis,
\begin{equation}H = \sum_{R_i,R_j} J(R_i-R_j)a^\dagger_{R_i}a_{R_j},\end{equation}where the function $J(R_i-R_j)$ describes how the tunneling matrix elements depend on the distance between the lattice sites. It is important to note that the Wannier functions are well localized and decay exponentially on larger distances \cite{Bloch2008}, so the dominant contribution is given by the hopping between adjacent lattice sites for sufficiently deep lattices. Then, the strength of this nearest-neighbor hopping is simply given by $J$.

\section{The Bose-Hubbard model}

Let us now look at the effect of interactions. Here, we assume that the interaction is weaker than the separation to the first excited Bloch band, i.e., we can still reduce our analysis to the lowest band. Additionally, we restrict the effect the effect of interactions to occur within the same lattice site, this is justified if the lattice is deep enough the Wannier functions decay rapidly. Then, we can approximate the optical lattice potential by a harmonic confinement and the Wannier function reduces to the ground state wave function of the harmonic oscillator,
\begin{equation}w(x) \approx \frac{1}{\sqrt{\sqrt{\pi}a_0}}e^{-x^2/(2a_0^2)},\end{equation}with the oscillator length being given by
\begin{equation}a_0 = \left(\frac{1}{4m^2V_0E_r}\right)^{1/4}.\end{equation}The matrix element for the contact interaction on a lattice site is then given by \cite{Bloch2008}
\begin{equation}U = g \int\mathrm{d}r |w(r)|^4 = \sqrt{\frac{8}{\pi}}k_L a_0 E_r \left(\frac{V_0}{E_r}\right)^{3/4}.\end{equation}Note that this interaction occurs for each pairwise combination of two-particles, so for $n_i$ particles on lattice site $i$, this interaction has to be weighted by a factor of $n_i(n_i-1)$.

Finally, we work in the grand-canonical ensemble, so there will be a chemical potential $\mu$ associated with the creation of a particle. This chemical potential can be site-dependent, in particular if we take the shape of the (typically harmonic) trapping potential into account. Using the localized Wannier states in the lowest Bloch band, the many-body Hamiltonian then takes the form \cite{Jaksch1998}
\begin{equation}H = -J \sum_{\langle ij\rangle}\left(a_ia_j^\dagger + \mathrm{h.c.}\right) + U\sum_i n_i(n_i-1) - \mu \sum_i n_i.\end{equation}This model is called the Hubbard model and is one of the most studied models of condensed matter physics. Here, we have found a way to quantum simulate its bosonic variant with ultracold atoms.

Let us now turn to the ground state phase diagram for the Bose-Hubbard model. For $J=0$, the Hamiltonian is classical and simplifies to a sum of on-site problems. Depending on the parameter $\mu/U$, there are different possible integer fillings per site, with phases having filling factor $n$ satisfying the relation
\begin{equation}
n-1\leq \frac{\mu}{U} \leq n.\end{equation}
If $J$ is nonzero but small, we can do perturbation theory in $J$. Then, in second order, we will create virtual particle or hole excitations. These excitations can move around, but essentially remain confined to the site where they originated. Consequently, this regime of the Bose-Hubbard model is an insulating phase (a Mott insulator).

On the other hand, in the limit of large $J$, the bosons will simply accumulate at the bottom of the single-particle band and form a Bose-Einstein condensate there. Therefore, there has to be a quantum phase transition in between separating these two regions.

\clearpage

To understand the behavior close to this insulator-superfluid transition, we make use of a variational ansatz that accounts for small fluctuations around the Mott insulator with a lattice filling $n^*$, taking the form of a product wave function \cite{Krauth1992},
\begin{equation}|\psi\rangle = \prod_i |\psi_0\rangle_i.\end{equation}In terms of the localized states, the on-site wave function takes the form
\begin{equation}|\psi_0\rangle = \sqrt{1-\varepsilon}|n^*\rangle + \sqrt{\varepsilon}|n^*\pm 1\rangle,\end{equation}where $\varepsilon$ is a variational parameter. As the wavefunction treats fluctuations increasing or decreasing the particle number equally, this ansatz is only correct when the average density satisfies $\langle n \rangle = n^*$, i.e., at the center of the Mott phases given by $\mu/U=n^*-1/2$. Then, the expectation value of the hopping reduces to
\begin{eqnarray}\langle \psi|a_ia_j^\dagger|\psi\rangle = (\langle \psi_0|a|\psi_0\rangle)^2 &=& (\langle \psi_0 |\sqrt{1-2\varepsilon}\sqrt{n^*}|n^*-1\rangle + \sqrt{\varepsilon}\sqrt{n^*+1}|n^*\rangle)^2\nonumber\\ &=& \varepsilon(1-2\varepsilon)(\sqrt{n^*}+\sqrt{n^*+1})^2.\end{eqnarray}For a lattice with $z$ nearest neighbors the total variational energy per lattice site is given by
\begin{equation}\langle H\rangle/N = -Jz\varepsilon(1-2\varepsilon)(\sqrt{n^*}+\sqrt{n^*+1})^2+U/2[2\varepsilon+n^*(n^*-1)] -\mu n^*.\end{equation}After minimization with respect to $\varepsilon$, we find a critical value $U_c$, at which $\varepsilon$ vanishes (and remains exactly zero for larger values of $U$), given by
\begin{equation}\frac{U_c}{zJ} = (\sqrt{n^*}+\sqrt{n^*+1})^2.\end{equation}This marks the quantum phase transition between the Mott insulator and the superfluid. For $n^*=1$, we find $U_c/zJ = 3+\sqrt{2} = 5.828$. For a three-dimensional square lattice ($z=6$), quantum Monte-Carlo results predict $U_c/zJ = 4.89$ \cite{Capogrosso-Sansone2007}, i.e., our simple variational ansatz predicts a value that is only 20\% too large. The overestimation of the superfluid phase is also quite natural: our variational wave function neglects any correlations between the sites and thus corresponds to a mean-field approximation \cite{Rokhsar1991}. Mean-field treatments are well-known to overestimate ordered phases such as superfluid as the order can be destroyed by the quantum fluctuation around the mean field that are being neglected. In lower dimensions, quantum fluctuations are even more important and the performance of mean-field theories becomes worse.

To map out the shape of the phase diagram in more detail, let us investigate what happens at small $J$ near the boundary between two Mott phases. There, only the two states $|n^*\rangle$ and $|n^*+1\rangle$ will be relevant. We use an ansatz of the form
\begin{equation}|\psi_0\rangle = \sqrt{1-\alpha}|n^*\rangle + \sqrt{\alpha}|n^*+1\rangle.\end{equation}Depending on our choice of $\mu/U$, either $\alpha$ or $1-\alpha$ will be a small quantity for small enough values of $zJ/U$. In the latter case, we find the lower branch of the Mott lobe as
\begin{equation}\frac{\mu_c}{U} = n-1 + \frac{2nzJ}{U},\end{equation}while the other case results in the upper branch given by
\begin{equation}\frac{\mu_c}{U} = n - \frac{2(n-1)zJ}{U}.\end{equation}The entire phase diagram can then be constructed by interpolation between these known limits.

\begin{figure}[t]

\includegraphics[width=12cm]{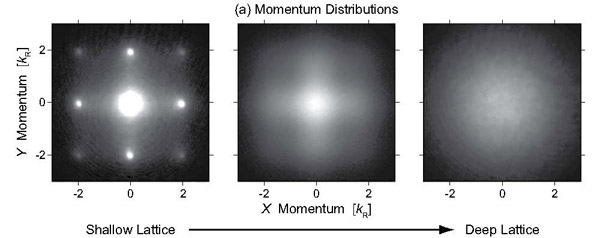}

\caption{Time-of-flight images of a two-dimensional Bose gas. For shallow lattices, the pronounced diffraction peaks indicate that the gas is in a superfluid phase. Increasing the lattice depth leads to stronger interactions and eventually results in the formation of a Mott insulator (taken from \cite{Spielman2007}).}

\end{figure}

Experimentally, the Mott transition can be probed in a time-of-flight experiment. For this, the trapping potentials are switched off and the atomic cloud is allowed to expand in free space. Then, an absorption image of the expanded cloud will correspond to the initial \textit{momentum} distribution of the atoms. In the superfluid, we expect a pronounced peak at $k=0$, which gets repeated when going to a higher-order Brillouin zones. In the Mott insulator, however, there is no preferred momentum so we expect a very broad distribution without any distinct features, see Fig. 1. The first experimental quantum simulation of the Mott insulator to superfluid transition were carried out in 2002 by Greiner et al., where they found a critical value of $U/zJ\approx 6$ \cite{Greiner2002}.

\section{Fermi-Hubbard model}

While the bosonic variant of the Hubbard model is well understood, this is not the case for the fermionic version, especially since the sign problem makes quantum Monte-Carlo calculations prohibitively hard. Using ultracold atoms, quantum simulation of the Fermi-Hubbard model can be achieved in a very analogous manner. However, since fermions cannot interact via an on-site interaction due to the Pauli principle, it is necessary to include a hyperfine spin degree as well so that fermions in different spin states can undergo a contact interaction, which is also characterized by its $s$-wave scattering length. Including spin, the Fermi-Hubbard model reads
\begin{equation}H = -J \sum_{\langle ij\rangle\sigma}\left(a_{i,\sigma}a_{j,\sigma}^\dagger + \mathrm{h.c.}\right) + U\sum_i n_{i\uparrow}n_{i\downarrow} - \mu \sum_{i\sigma} n_{i\sigma}.\end{equation}
Despite its hardness, several things can be observed about the Fermi-Hubbard model. In the case of half filling and in the limit of $J\ll U$, we can employ perturbation theory in $J$. At $J=0$, all the sites are filled with a single fermion, and there are no double occupancies anywhere in the system. However, for finite $J$, we can virtually create holes $|0\rangle$ and doublons $|D\rangle$. Due to the Pauli principle, this can only happen when the spins are opposite on the neighboring sites. In lowest order, the following terms are relevant:
\begin{eqnarray}\langle \uparrow \downarrow|a_ia_{i+1}^\dagger|D0\rangle\langle D0|a_i^\dagger a_{i+1}|\uparrow \downarrow\rangle = -1\\\langle \uparrow \downarrow|a_ia_{i+1}^\dagger|D0\rangle\langle D0|a_i^\dagger a_{i+1}|\downarrow \uparrow\rangle = 1.\end{eqnarray}Then, the model can be represented by local spin-1/2 degrees of freedom and reduces to the antiferromagnetic Heisenberg model,
\begin{equation}H=\frac{zJ^2}{U} \sum_{\langle ij\rangle} \textbf{S}_i\textbf{S}_j.\end{equation}On the two-dimensional square lattice, there is no sign problem, and quantum Monte-Carlo results have found the existence of an antiferromagnetically ordered ground state \cite{Sandvik1997}. Away from half-filling, the situation is much more subtle, however, there is the striking possibility of the appearance of an unconventional $d$-wave superconducting phase, which might explain the mystery of high-temperature superconductivity in copper oxides \cite{Zhang1988}.

The status of experiments with ultracold fermions in optical lattices is not as far developed as for bosons, as the cooling process is more challenging. In addition, the energy scales associated with antiferromagnetic order or even $d$-wave superconductivity are a fraction of $J^2/U$, which requires to cool down to extremely low temperatures. In 2008, two groups achieved the realization of a Mott insulating phase with fermions \cite{Jordens2008,Schneider2008}.

\bibliographystyle{apsrev} 
\bibliography{/home/itp/weimer/hendrik}

\end{document}


\title{Chapter 1: Exact Diagonalization}

\maketitle

\section{The many-body problem}

A common theme in all branches of quantum physics is to identify the eigenstates $|\phi_i\rangle$ of a Hamiltonian $H$, and the respective eigenenergies $E_i$. Once equipped with the eigenstates, we can relatively easily calculate many interesting aspects such as the time evolution, or its thermal properties, according to the relation

\begin{equation}
\rho = Z^{-1}\exp(-\beta H) = Z^{-1} \sum\limits_i \exp(-\beta E_i)|\phi_i\rangle\langle\phi_i|,
\end{equation}
where $Z=\mathrm{Tr}\{\exp(-\beta H)\}$,
which gives us the density operator $\rho$ at the inverse temperature $\beta$. A special case occurs at (or close to) zero temperature, as then only the ground state $|\phi_g\rangle$ will provide the dominant contribution. Nevertheless, the system can still undergo phase transitions if the ground state energy $E_g$ itself exhibits non-analytic behavior. Consequently, finding the ground state is often a crucial step to establish a thorough understanding of a quantum system.

In the following, we will mainly focus on discussing ground state properties of many-body systems containing only few local degrees of freedom. A prominent class of such many-body models are spin $1/2$ models, which consist of two-level systems localized at the sites of some lattice. As a specific example, let us turn to the to the one-dimensional transverse field Ising model, whose Hamiltonian is given by

\begin{equation}H = g\sum\limits_{i=1}^N  \sigma_x^{(i)} - \sum\limits_{i=1}^{N-1}\sigma_z^{(i)}\sigma_z^{(i+1)}.
\end{equation}

Here, we have expressed the spins with the help of Pauli matrices, while $g$ can be interpreted as the strength of a field transverse to the quantization axis of the spins, and all energies are measured in the strength of the interaction between two neighboring spins. While this model is actually exactly solvable in the limit $N \to \infty$ \cite{Sachdev1999}, we will use it to illustrate various approaches to many-body problems.

Without going into specific details, let us consider the two possible limits of the transverse Ising model. For $g\to\infty$, we can ignore the interaction, and the problem reduces to a single spin in a magnetic field. Consequently, the ground state is given by
\begin{equation}|\phi_g\rangle = \prod\limits_{i=1}^N \frac{1}{\sqrt{2}}(|\uparrow\rangle_i-|\downarrow\rangle_i).
\end{equation}
In the limit of zero external field, $g\to 0$, we just have to minimize the interaction energy, which leads to two distinct fully polarized ground states, $|\phi_g^{(1)}\rangle = |\downarrow\downarrow\downarrow\cdots\rangle$ and $|\phi_g^{(2)}\rangle = |\uparrow\uparrow\uparrow\cdots\rangle$. Note that the Hamiltonian does not distinguish between up and down spins (a so-called $\mathbb{Z}_2$ symmetry), but the two possible ground states do. This symmetry breaking is a manifestation of the system being in two different quantum phases for $g \gg 1$ (ferromagnet) and $g \ll 1$ (paramagnet). From the exact solution, it is known that the quantum phase transition occurs at a critical transverse field of $g_c = 1$.

\section{Exact and not-so-exact diagonalization}

The term ``exact diagonalization'' is often used in a slightly misleading manner. In general, to find the eigenvalues of a $d$-dimensional Hamiltonian, one has to find the roots to the characteristic polynomial of degree $d$, for which in general no exact solution can be found for $d>4$. Of course, we can still hope to numerically approximate the eigenvalues to an arbitrary degree, but the fact that we have to work with computers operating with fixed precision numbers makes this endeavour substantially more complicated.

Keeping that aside, finding the eigenvalues by solving the characteristic polynomial is a bad idea, as finding roots of high-degree polynomials is a numerically tricky task \cite{Press1992}. In fact, one of the most powerful methods for finding the roots of such a polynomial is to generate a matrix that has the same characteristic polynomial and find its eigenvalues using a different algorithm! A much better strategy is to find a unitary (or orthogonal, if all matrix elements of the Hamiltonian are real) transformation that makes the Hamiltonian diagonal, i.e.,

\begin{equation} H \to U^{\dagger} H U\end{equation}

The general strategy is to construct the matrix $U$ in an iterative way,

\begin{equation} H \to U_1^\dagger H U_1 \to U_2^\dagger U_1^\dagger H U_1 U_2 \to \cdots
\end{equation}

until the matrix becomes diagonal. The columns of $U = U_1 U_2 U_3 \cdots$ then contains the eigenvectors of $H$. There are many different algorithms for actually performing the diagonalization, and it is usually a good idea to resort to existing libraries (such as LAPACK) for this task. However, if we are only interested in finding the low-energy eigenvalues of a particular Hamiltonian, we can make a few simplifications that will allow for much faster computations.

\section{The power method}

Initially, we pick a random state $|\phi_0\rangle$ which has a very small but finite overlap with the ground state, i.e., $\langle \phi_0|\phi_g\rangle \ne 0$. Then, we repeatedly multiply the Hamiltonian with this initial state and normalize the result,

\begin{equation}|\phi_{n+1}\rangle = \mathcal{N} H |\phi_n\rangle,\end{equation}
where $\mathcal{N}$ is the normalization operation. This method will eventually converge to the eigenvector with the largest absolute eigenvalue, so by subtracting a constant energy from the Hamiltonian, we can always ensure that this will be the ground state. The key advantage is that the matrix-vector multiplications occuring at each iteration can be implemented very fast: in most cases (as for the transverse Ising model), for each column the number of nonzero entries in such a sparse Hamiltonian are much smaller than the dimension of the Hilbert space (here: $N$ vs. $2^N$).

\section{Lanczos algorithm}

The power method can be readily improved by using not only a single state during each iteration, but employing a larger set of states which will be extended until convergence is reached. This procedure is known as Lanczos algorithm and is implemented as follows \cite{Dagotto1994}:
\begin{enumerate}
\item  Pick a random state $|\phi_0\rangle$ as in the Power method
\item  Construct a second state $|\phi_1\rangle$ according to
\begin{equation}|\phi_1\rangle = H|\phi_0\rangle - \frac{\langle\phi_0|H|\phi_0\rangle}{\langle \phi_0|\phi_0\rangle} |\phi_0\rangle\end{equation}\item  Starting with $n=2$, recursively construct an orthogonal set of states given by
\begin{equation}|\phi_{n+1}\rangle = H|\phi_n\rangle - a_n |\phi_n\rangle - b_n^2|\phi_{n-1}\rangle,\end{equation} where the coefficients $a_n$ and $b_n$ are given by
\begin{equation} a_n = \frac{\langle\phi_n|H|\phi_n\rangle}{\langle \phi_n|\phi_n\rangle}\;\;\;b_n^2 = \frac{\langle\phi_n|\phi_n\rangle}{\langle \phi_{n-1}|\phi_{n-1}\rangle}.\end{equation}\item  Diagonalize the matrix given by
\begin{equation} H = \left(\begin{array}{ccccc}a_0 & b_1 & 0 & 0 & \cdots\\
b_1 & a_1 & b_2 & 0 & \cdots\\
0 & b_2 & a_2 & b_3 & \cdots\\
0 & 0 & b_3 & a_3 & \cdots\\
\vdots & \vdots & \vdots & \vdots & \ddots\end{array}\right).\end{equation}
\item  If the ground state energy has not converged to the desired accuracy, proceed at step 3 by increasing $n$ by one.
\end{enumerate}


\begin{figure}[b]
  \includegraphics{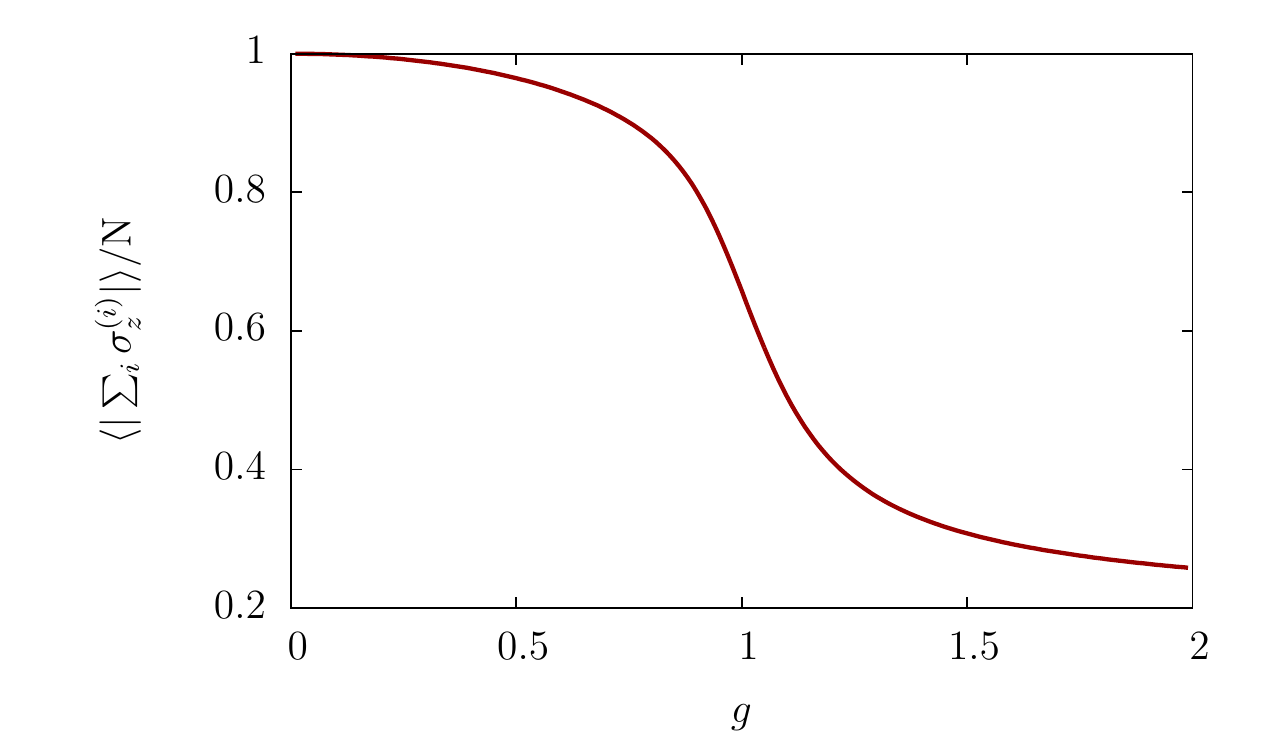}
  \caption{Spontaneous magnetization of the ferromagnetic 1D Ising model in a transverse field for 18 spins computed using the Lanczos algorithm.}
\end{figure}

While the exact ground state is only reached when $n$ is equal to the dimension of the Hilbert space, the remarkable feature of the Lanczos algorithm is that typically only a few hundred iterations are necessary. However, there is one important caveat: due to finite-precision arithmetic, the orthogonality between the states $|\phi_n\rangle$ is quickly lost. For practical applications, it is therefore advisable to use an improved algorithm that re-orthogonalizes the states.

\section{Some remarks on complexity}

Let us briefly consider the difficulty inherent in exact diagonalization studies. On each lattice site, we have 2 degrees of freedom, refering to a spin pointing either up or down. However, for $N$ spins, we have $2^N$ possible spin configuration. The consequences of this exponential scaling cannot be underestimated: if we want to store the vector corresponding to the ground state in a computer using double precision, we would need 8 TB of memory for $N=40$, and for $N=300$, the number of basis states already exceeds the number of atoms in the universe! From these considerations, we see that exact diagonalization can only work in the limit of small system sizes.

\bibliographystyle{apsrev} 
\bibliography{/home/itp/weimer/hendrik}


\title{Chapter 2: Quantum Monte-Carlo}

\maketitle

\section{Quantum-classical mapping}

We have seen in the previous chapter that exact diagonalization is an impossible task for more than a few particles. In quantum Monte-Carlo simulations, the goal is to avoid considering the full Hilbert space, but randomly sample the most relevant degrees of freedom and try to extract the quantities of interest such as the magnetization by averaging over a stochastic process. To pursue this goal, we first need to establish a framework in which we can interpet quantum-mechanical observables in terms of (classical) probabilities. This process is called a "quantum-classical mapping" and allows us to reformulate quantum many-body problems in terms of models from classical statistical mechanic, albeit in higher dimensions.

Suppose we wish to calculate the partition function $Z=\mathrm{Tr}\{\exp(-\beta H)\}$ of the transverse Ising model, as knowledge of the partition function allows us to calculate all thermodynamic quanties that may be of interest. Specifically, we have
\begin{align}Z &= \mathrm{Tr}\left\{\exp\left[-\beta\left(g\sum\limits_{i}\sigma_x^{(i)}-\sum\limits_i\sigma_z^{(i)}\sigma_z^{(i+1)}\right)\right]\right\}\\&=\mathrm{Tr}\left\{\exp\left(-\frac{\beta g}{N_y}\sum\limits_{i}\sigma_x^{(i)}+\frac{\beta}{N_y}\sum\limits_i\sigma_z^{(i)}\sigma_z^{(i+1)}\right)^{N_y}\right\}\\
& = \lim\limits_{N_y\to\infty}\mathrm{Tr}\left\{\left[\exp\left(-\frac{\beta g}{N_y}\sum\limits_{i}\sigma_x^{(i)}\right)\exp\left(\frac{\beta}{N_y}\sum\limits_i\sigma_z^{(i)}\sigma_z^{(i+1)}\right)\right]^{N_y}\right\},\end{align}where in the last line we have used the Suzuki-Trotter expansion
\begin{equation}\exp\left[\frac{1}{N}(A+B)\right] = \exp\left(\frac{A}{N}\right)\exp\left(\frac{B}{N}\right)+O(1/N^2),\end{equation}which can be proved using the Baker-Campbell-Hausdorff formula. Using the same expansion, we can replace the exponential of the product by a product of the exponentials, i.e.,
\begin{equation}
Z=\lim\limits_{N_y\to\infty}\mathrm{Tr}\left\{\exp\left(-\frac{\beta g}{N_y}\sum\limits_{i}\sigma_x^{(i)}\right)^{N_y}\exp\left(\frac{\beta}{N_y}\sum\limits_i\sigma_z^{(i)}\sigma_z^{(i+1)}\right)^{N_y}\right\}.
\label{eq:Z}
\end{equation}
The exponentiation to the power of $N_y$ can be written as a product, and we can insert $N_y-1$ identity operators according to
\begin{equation}A^{N_y} = \prod\limits_{i=1}^{N_y}A = A|i_1\rangle\langle i_1| A |i_2\rangle\langle i_2|A \cdots A|i_{N_y-1}\rangle\langle i_{N_y-1}|A.\end{equation}Let us know look more closely at the product involving the spin-flip operators $\sigma_x$, where we will encounter terms of the form 
\begin{equation}\langle i_j|\exp\left(a\sigma_x^{(i)}\right)|i_{j+1}\rangle = \langle i_j|\left[\cosh(a) + \sigma_x^{(i)}\sinh(a)\right]|i_{j+1}\rangle.
\end{equation}
The crucial part of the quantum-classical mapping is to interpret the partition function of the one-dimensional chain containing $N$ spins as the partition function of a corresponding two-dimensional spin model containing $N \times N_y$ spins \cite{Batrouni2011}. In this interpretation, we can rewrite the spin-flip operators in terms of an Ising interaction in the $y$ direction (plus a constant energy shift),
\begin{equation}\langle i_j|\left[\cosh(a) + \sigma_x^{(i)}\sinh(a)\right]|i_{j+1}\rangle = \frac{1}{2}[\exp(-a)  \langle i_j |\sigma_z^{(i,j)}\sigma_z^{(i,j+1)}|i_{j+1}\rangle +\exp(a)].\end{equation}We now want to cast these terms back into an exponential form,
\begin{equation}\frac{1}{2}[\exp(-a)  \langle i_j |\sigma_z^{(i,j)}\sigma_z^{(i,j+1)}|i_{j+1}\rangle +\exp(a)] = \Lambda\exp\left(\gamma \sigma_z^{(i,j)}\sigma_z^{(i,j+1)}\right)\end{equation}where we find for the coefficents $\Lambda$ and $\gamma$
\begin{align}\Lambda &= \sqrt{\sinh(a)\cosh(a)}\\
\gamma & = -\frac{1}{2}\log\tanh(a).\end{align}We can now insert this expression back into the partition function Eq.~(\ref{eq:Z}) and carry out the $N_y$ multiplications. The boundary conditions for the Ising interaction along the $N_y$ direction are fixed by the final trace operation; as the trace is implemented by multiplying $\langle i_{N_y}|$ from the left and $|i_{N_y}\rangle$ from the right, we have periodic boundary conditions. In total, we obtain
\begin{equation}Z = \Lambda^{NN_y}\mathrm{Tr}\left\{\exp\left(\gamma\sum\limits_{i=1}^N\sum\limits_{j=1}^{N_y}\sigma_z^{(i,j)}\sigma_z^{(i,j+1)} + \frac{\beta}{N_y}\sum\limits_{i=1}^N\sum\limits_{j=1}^{N_y}\sigma_z^{(i,j)}\sigma_z^{(i+1,j)}\right)\right\}.\end{equation}The constant prefactor in the partition function is irrelevant as it will drop out when calculating thermodynamic observables. Consequently, we can identify a corresponding two-dimensional classical Ising model with anisotropic interactions which reproduces the same thermodynamics as the one-dimensional quantum Ising model in a transverse field. The Hamiltonian for the classical model is given by
\begin{equation}H_{cl} = -\frac{N_y\gamma}{\beta}\sum\limits_{i=1}^N\sum\limits_{j=1}^{N_y}\sigma_z^{(i,j)}\sigma_z^{(i,j+1)} - \sum\limits_{i=1}^N\sum\limits_{j=1}^{N_y}\sigma_z^{(i,j)}\sigma_z^{(i+1,j)}.\end{equation}Note, however, that the classical temperature $\beta_{cl} = \beta/N_y$ is different from the quantum temperature $\beta$. Nevertheless, we can now proceed to calculate the thermodynamic properties of the quantum model by performing classical Monte-Carlo simulations.

\section{Metropolis algorithm}

When trying to evaluate thermodynamic observables for a classical spin model, we still find ourselves in considerable difficulties as also the classical configuration space grows exponentially with system size. However, we are not really interested in a solution that incorporates all microscopic details, but rather we want to obtain information about macroscopic observables. So, we will be fine with any microscopic description of the model of interest, as long as it gets the macroscopic statistics right. Here, the goal is to find a microscopic description which can be efficiently (i.e., using resources that only grow polynomially with system size) implemented on a computer.

The most famous method for the Monte-Carlo simulation of statistical mechanics models is the Metropolis algorithm \cite{Metropolis1953}. Let us first state the basic steps of the algorithm for the Ising model and then analyze it in more detail. 
\begin{enumerate}
\item Pick an arbitrary initial state (e.g., all spins polarized) and compute its energy $E$.
\item Flip a random spin and calculate the energy of the new configuration $E'$
\item If $E'<;E$, always accept the new configuration.
\item If $E'>E$, accept the new configuration with probability $\exp(-\beta[E'-E])$.
\item Continue at step 2 until the macroscopic observables (averaged over a fixed number of steps) are equilibrated.
\end{enumerate}


\begin{figure}[ht]
\begin{tabular}{c@{\hspace{1cm}}c}
  \includegraphics{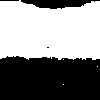} & \includegraphics{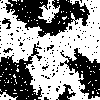}
\end{tabular}
\caption{Two-dimensional classical Ising model at $\beta=1$ (left) and at $\beta=\beta_c=\log(1+\sqrt{2})/2$ (right) (taken from \cite{Altervista2007}).}
\end{figure}

To evaluate the algorithm, let us consider two configurations $x$ and $x'$ with energies $E$ and $E'$,respectively. The probalities for these configurations are denoted by $p(x)$, and $p(x')$. Let us assume that $E<E'$, so the transition probability satisfy $p(x'\to x)=1$ and $p(x'\to x)=\exp(-\beta[E'-E])$ for the inverse process. In equilibrium, the system will satisfy detailed balance (absence of currents), i.e,
\begin{equation}p(x)p(x\to x') = p(x')p(x'\to x),\end{equation}or equivalently
\begin{equation}\frac{p(x')}{p(x)} = \frac{p(x\to x')}{p(x'\to x)} = \exp(-\beta[E'-E]),\end{equation}which reproduces the Boltzmann distribution and thus gives rise to the correct thermodynamic behavior. The convergence of the Metropolis algorithm can be improved by also including unphysical processes that flip a large domain of spins at once (so-called "cluster updates") \cite{Krauth2004}, as constructing these processes from individual spin flips will consume a lot of time. Fig. 1 shows results of Monte-Carlo simulations for two different values of $\beta$ for the isotropic two-dimensional Ising model.

\clearpage

\section{From classical to quantum phase transitions}

Let us now come back to the anisotropic Ising model that we obtained using the quantum-classical mapping. If we play around with $N_y$ and $\beta_{cl}$, we find that the phase transition between the paramagnet and the ferromagnet occurs on a critical line that is given by
\begin{equation}\sinh(2\beta_{cl}^c)\sinh\left(2\beta_{cl}^c \frac{N_y\gamma}{\beta}\right) = 1,\end{equation}where $\beta_{cl}^c$ is the critical inverse temperature \cite{Batrouni2011}. Remarkably, the condition $N_y \to \infty$ at a fixed classical temperature $\beta_c = \beta/N_y$ means that the corresponding quantum phase transition takes place at $\beta=\infty$, i.e., at zero temperature. At finite temperature, the extension in the $N_y$ direction will be finite and the classical model belongs to the universality class of the one-dimensional Ising model, which does not exhibit a phase transition. If we substitute the definitions for the coupling constants into the equation for the critical line, we obtain
\begin{equation}\frac{\sinh(2\beta_{cl}^c)}{\sinh(2 g_c\beta_{cl}^c)},\end{equation}which immediately gives us the critical transverse field $g_c = 1$, in agreement with the exact solution of the quantum model. From our knowledge of the classical 2D Ising model, we can also immediate extract the scaling of observables close to the critical point, e.g., for the magnetization
\begin{equation}m = m_0(g-g_c)^{1/8},\end{equation}as the critical exponents for the 1D quantum model are identical to the ones of the classical 2D model.

\section{The sign problem}

While quantum Monte-Carlo methods are indeed very powerful for tackling a large variety of many-body problems, they are also constrained by inherent limitations. One possible problem we can run into is that the partition function of the classical model cannot be computed efficiently. For example, consider the classical Hamiltonian
\begin{equation}H = \sum\limits_{\langle i,j\rangle}J_{ij}\sigma_z^{(i)}\sigma_z^{(j)}\end{equation}on a three-dimensional cubic lattice, where the nearest-neighbor interactions $J_{ij}$ are randomly chosen from the values $\{-1,0,+1\}$. This model exhibits glassy behavior and finding its ground state is known to be NP-complete \cite{Barahona1982}, i.e., it is widely believed that there is no efficient way to simulate this model. On the other hand, we can safely assume that precisely due to the glassy properties that makes the model hard to study, the ground state is irrelevant as a physical state as the system will take exponentially long to reach it even if we put it into heat bath at zero temperature. A more subtle issue can arise as the result of the quantum-classical mapping, which is known as the ``sign problem''.

To be explicit, let us assume that during our quantum-classical mapping procedure, we encounter the following term in the Hamiltonian:
\begin{equation}H = J \sigma_+^{(i)}\sigma_-^{(i+1)} + \mathrm{h.c.},\end{equation}where we have used the spin-flip operators $\sigma_+ = |\uparrow\rangle\langle\downarrow|$ and $\sigma_- = \sigma_+^\dagger = |\downarrow\rangle\langle\uparrow|$.
Within the quantum-classical mapping, we will encounter terms of the form
\begin{equation}\langle i_j|\exp\left(a \sigma_+^{(i)}\sigma_-^{(i+1)}\right)|i_{j+1}\rangle = \frac{1}{2}[\cosh(a)+1 + (\cosh(a)-1)\sigma_z^{(i)}\sigma_z^{(i+1)} + \sinh(a)\sigma_+^{i}\sigma_-^{i+1}].\end{equation}The constant term and the Ising interaction are unproblematic and can be cast into the classical partition function as in the case of the transverse Ising model. Whether we can do the same for the spin-flip term (which will ultimately result in a four-body interaction for the classical Ising spins), depends on the sign of $J$. For the ferromagnetic case, $J<0$, the coefficient $a = -\beta J/N$ will be positive and we can turn this interaction into a term of the form $\Lambda \exp(-\beta_{cl}J_{cl}H_{cl})$ with $\Lambda$ being positive so the interpretation in terms of a classical partition function remains valid. On the other hand, for an antiferromagnetic interaction with $J>0$, we find $\Lambda$ to be negative and therefore we do not obtain the equivalent of a classical partition function. In some cases, we can work around this by using a unitary transformation that maps onto a model that does not exhibit this problem. For instance on a bipartite lattice, we can fix this problem by flipping spins on one of the sublattices
by transforming $\sigma_z \to -\sigma_z$, turning the antiferromagnetic interaction into a ferromagnetic one. However, this does not work in general, nevertheless there is still a way to perform classical Monte-Carlo sampling. When sampling over our classical configuration space, the problematic contributions correspond to negative probabilities, i.e., when we compute the value of an observable $A$,
\begin{equation}\langle A\rangle = \frac{\mathrm{Tr}\{A\exp(-\beta H)\}}{\mathrm{Tr}\{\exp(-\beta H)\}}=\frac{\sum\limits_iA_ip_i}{\sum\limits_i p_i},\end{equation}we find some of the $p_i$ to be negative. This can be resolved by taking the absolute value of these probabilities and dividing by the expectation value of the sign, i.e.,
\begin{equation}\langle A\rangle = \frac{\sum\limits_i A_i \sgn p_i |p_i|/\sum_i|p_i|}{\sum\limits_i\sgn p_i |p_i|/\sum_i |p_i|} \equiv \frac{\langle As\rangle_{|p|}}{\langle s\rangle_{|p|}}.\end{equation}However, this modification comes at a hefty price: as the uncertainty of the denominator will typically increase exponentially with the size of the system \cite{Troyer2005}, we will no longer have an efficient algorithm.

Despite these difficulties, it may sometimes still be more favorable to perform Monte-Carlo simulation than performing exact diagonalization, but this certainly depends on details of the model at hand. Finally, the sign problem is not limited to models of the type outlined above. As a rough guidance, frustrated spin models and fermionic models tend to have a sign problem, while bosonic models tend to be free of it.

\bibliographystyle{apsrev} 
\bibliography{/home/itp/weimer/hendrik}


\title{Chapter 3: Density-Matrix Renormalization Group}

\maketitle

\section{Variational method}

While quantum Monte-Carlo methods provide very useful tools to study many-body systems, we have seen that the sign problem consittutes a significant challenge. In addition, the required quantum-classical mapping makes it difficult to implement direct computer algorithms taking only the Hamiltonian as their input. On the other hand, there is a conceptually simple but quite powerful method to approximate the ground state of arbitrary quantum systems which is known as the variational method. In a nutshell, one writes down an ansatz for the ground state as a trial wave function, which contains one or more variational parameters that are chosen such that the minimize the energy of the system.

As a specific example, let us study the Hamiltonian for the Helium atom,
\begin{equation}H = \sum\limits_{i=1,2}\frac{p_i^2}{2} - \sum\limits_{i=1,2} \frac{Z}{r_i} + \frac{1}{|\mathbf{r}_1-\mathbf{r}_2|},\end{equation}where we have used atomic units, $Z=2$ denotes the charge of the nucleus, and the positions $\mathbf{r}_i$ and momenta $\mathbf{p}_i$ satisfy the usual commutation relations $[r_{i,\alpha},p_{j,\beta}] = i\delta_{ij}\delta_{\alpha\beta}$ for the components $\alpha,\beta=x,y,z$. Physically speaking, it makes sense that the interaction between the electrons shield the Coulomb potential of the nucleus, so we use as an ansatz a product wave function of the form given in position space by \cite{Schwabl2010}
\begin{equation}\psi(\mathbf{r}_1,\mathbf{r}_2) = \frac{\tilde{Z}^3}{\pi}\exp(-\tilde{Z}[r_1+r_2]),\end{equation}where the variational parameter $\tilde{Z}$ accounts for the shielding of the core. Computing the expectation value of the Hamiltonian with respect to this trial wave function yields
\begin{equation}\langle H \rangle = \tilde{Z}^2 - 4\tilde{Z} + \frac{5}{8}\tilde{Z}.\end{equation}This function has a minimum at the effective charge $\tilde{Z}=27/16=1.6875$ and leads to a ground state energy of $E_g = -729/256\approx -2.848$, which is within 2\% accuracy of the experimentally measured value of $E_g = -2.903$ \cite{Sansonetti2005}.


\section{Reduced density matrices}

As the name implies, the Density-Matrix Renormalization Group (DMRG) method does not operate on pure quantum states, but on density matrices, which were originally discussed in the context of open quantum systems. Generally, we can think of a system decribed by a density matrix $\rho$ as a statistical mixture of pure quantum states $|\psi_i\rangle$, i.e.,
\begin{equation} \rho = \sum\limits_i p_i |\psi_i\rangle\langle\psi_i|,\end{equation}where $p_i$ can be interpreted as the probability to find the system in the pure quantum state $|\psi_i\rangle$. Density matrices have unit trace and are positive-semidefinite, and their time evolution is governed by the von Neumann equation,
\begin{equation}\frac{\mathrm{d}}{\mathrm{dt}}\rho = -\frac{i}{\hbar} [H, \rho].\end{equation}Density matrices play an important role when looking at a smaller subsystem embedded in an environment. Consider a bipartite system composed of the subsystems $A$ and $B$, with pure states being represented by
\begin{equation}|\psi\rangle = \sum\limits_{ij} c_{ij} |i\rangle_A|j\rangle_B \equiv \sum\limits_{ij} c_{ij}|ij\rangle.\end{equation}Then, we can define the reduced density matrix of $A$ as the partial trace over $B$, given by
\begin{equation}\rho_A = \mathrm{Tr}_B\{\rho\} = \sum\limits_{i,i'}\sum\limits_j \langle ij|\rho|i'j\rangle |i\rangle\langle i'|.\end{equation}

\section{The DMRG algorithm}

The key element of DMRG is to think of the many-body system of interest being composed of a system of size $l$, attached to an environment of the size $L-l$. Suppose we have already found a set of $D$ states $\{|\phi_S\rangle\}$, which allows us to give a good approximation to both the Hamiltonian and its ground state for this particular system size. Then, we can increase the system size to $l+1$ by the following procedure \cite{Schollwock2005}:
\begin{enumerate}
\item Form a new system $S'$ of size $l+1$ by combining the $D$ states describing the system with the full Hilbert space describing the additional site. For spin $1/2$, we have a new set of states according to $\{|\phi^S_l\uparrow\rangle,|\phi^S_l\downarrow\rangle\}$.
\item Build a superblock of size $2l+2$ by combining the enlarged system with the enlarged environment. The Hilbert space of this dimension will be $4D^2$ in the case of spins. If the Hamiltonian is reflection symmetric (i.e, left and right are identical), the basis states for the system and the environment will be the same.
\item Find the ground state $|\psi\rangle$ of the Hamiltonian of this superblock by exact diagonalization.
\item Compute the reduced density matrix for the enlarged system
\begin{equation}\rho_S' = \mathrm{Tr}_E'\{|\psi\rangle\langle\psi|\}.\end{equation}
\item Diagonalize the reduced density matrix and keep only the eigenvectors corresponding to the $D$ largest eigenvalues. This set of states $\{|\psi_{l+1}^S\rangle\}$ will serve as the input for the next iteration step.
\item Express the Hamiltonian for system size $l+1$ in this new basis.
\item Continue this iteration until sufficient convergence of observables (e.g, of the ground state energy) is obtained.
\end{enumerate}

This procedure is commonly referred to as ``infinite-system DMRG'' \cite{Schollwock2005}. In practice, one can often achieve significantly better results by accounting for finite size effects. This can be done in a relatively straightforward way by keeping track of the system and the environment separately. Once the infinite-system algorithm has reached the desired size, one can successively increase the system at the expense of the environment. When the environment reaches a minimum size, the procedure is reversed and the environment is increased at the expense of the system. Multiple sweeps of this kind can be performed until the ground state energy has converged to an even better value. During each step, we can also use the estimate for the ground state obtained during the previous step as the initial state of our exact diagonalization procedure, which will result in a drastically improved performance \cite{Schollwock2005}. In general, the error of the DMRG procedure will be given by the truncation error $\varepsilon$, which is the weight of all the states that have been dropped during the DMRG step. Typical values are $\varepsilon \sim 10^{-10}$ for $D\sim 100$.

As in the case of exact diagonalization, DMRG is not restricted to ground states, but due to the constraints imposed by the convergence of the exact diagonalization step, it works best for low-energy eigenvalues. It is also possible to compute the time evolution of a many-body system using DMRG, for instance by combining it with the Runge-Kutta scheme for the numerical integration of ordinary differential equations \cite{Feiguin2005}. This is done by using the DMRG procedure to converge towards four different target states,
\begin{align}
|k_1\rangle &= \tau H(t)|\psi(t)\rangle\\
|k_2\rangle &= \tau H(t+\tau/2)\left[|\psi(t)\rangle + \frac{1}{2}|k_1\rangle\right]\\
|k_3\rangle &= \tau H(t+\tau/2)\left[|\psi(t)\rangle + \frac{1}{2}|k_2\rangle\right]\\
|k_3\rangle &= \tau H(t+\tau)\left[|\psi(t)\rangle + |k_3\rangle\right].\end{align}The state at the time $t+\tau$ then follows from \cite{Press1992}
\begin{equation}|\psi(t+\tau)\rangle = \frac{1}{6}[|k_1\rangle + 2|k_2\rangle + 2|k_3\rangle + |k_4\rangle] + O(\tau^5).\end{equation}

\section{Matrix product states}

To understand the success of DMRG, it is instructive to look at the set of states that can be created using this method. It turns out that these are given by so-called ``matrix product states'' of the form
\begin{equation}
|\psi\rangle = \sum\limits_{\{i_j\}} \mathrm{Tr}\left\{\prod\limits_{k=1}^N A_k^{i_k}\right\}|i_1,i_2,\ldots\rangle,
\end{equation}where $A_k^{i_k}$ is a $D\times D$ matrix associated with every site $k$ and the corresponding basis vector $|i_k\rangle$. For translational invariant systems, the matrix elements will be identical for all sites, but each matrix will still act on its own (axillary) Hilbert space, i.e., we can express each $A_j$ as
\begin{equation}A_j = \sum\limits_{\alpha\beta}(A)_{\alpha\beta}|\alpha\rangle_j\langle\beta|_j,\end{equation}for some orthonormal basis $\{|\alpha\rangle\}$. During each DMRG step, we perform basis transforms of the form
\begin{equation}|m_l\rangle = \sum_{m_{l-1},\sigma_l} = (A_l)_{m_l,m_{l-1}}|m_{l-1}\rangle|\sigma_l\rangle,\end{equation}which tell us how to go from the block of size $l-1$ to size $l$ by including an additional site, with $\sigma_l = \uparrow,\downarrow$ for spin $1/2$ \cite{Schollwock2005}. Applying this construction recursively, we see that the basis states obtained within DMRG indeed belong to the class of matrix product states. The variational character of the DMRG algorithm becomes apparent when we consider the state of a superblock,
\begin{equation}|\psi\rangle = \sum\limits_{mn}^D \sum\limits_i \psi_{m\sigma_{l/2}\sigma_{l/2+1}n} (A_{l/2-1}\cdots A_M)_{m,(\sigma_1\cdots \sigma_M)} (A_{l/2+2}\cdots A_{N-M})_{n,(\sigma_{N-M}\cdots\sigma_N)}|i\rangle,\end{equation}where $M$ forms the largest block that can be solved exactly, i.e, $\dim(M) = D$. Then, we can interpret the coefficients $\psi_{m\sigma_{l/2}\sigma_{l/2+1}n}$as the variational parameters according to which the energy is being minimized. Further information about the relationship between DMRG and matrix product states can be found in a more recent review article \cite{Schollwock2011}.


\section{Entanglement and the area law}

Having seen that DMRG implements a variational method based upon matrix product states, it is important to pose the question how generic these states will be in terms of ground states of many-body Hamiltonians. A key concept in answering this question is entanglement, i.e., nonclassical correlations between different parts of the system. Consider the two-spin state
\begin{equation}|\psi\rangle = \frac{1}{\sqrt{2}}(|\uparrow\uparrow\rangle+|\downarrow\downarrow\rangle,\end{equation}which can be represented by the density matrix
\begin{equation}\rho = \left(\begin{array}{cccc}1/2 & 0 & 0 & 1/2\\0 & 0 & 0 & 0\\ 0 & 0 & 0 & 0\\1/2 & 0 & 0 & 1/2\end{array}\right).\end{equation}As can be seen from the diagonal elements, this state possesses classical correlations, but its off-diagonal contributions are what is crucial here. It is instructive to look at the reduced density matrix of a single spin,
\begin{equation}\rho_1 = \mathrm{Tr}_2\{\rho\} = \left(\begin{array}{cc}1/2 & 0\\0 & 1/2\end{array}\right).\end{equation}This state is proportional to the identity, so it is invariant under all unitary transformations, i.e., $U\rho U^\dagger=\rho$. Consequently, despite the two-spin state being pure, its local density matrix of a single spin is completely random and does not carry any information at all!

We can quantify this behavior using the von Neumann entropy, defined as
\begin{equation}S = -\mathrm{Tr}\{\rho\log\rho\},\end{equation}which vanishes for pure states and reaches its maximum $\log d$, with $d$ being the Hilbert space dimension, for the ``maximally mixed state'',
\begin{equation}\rho = \left(\begin{array}{ccc}1/d \\ & 1/d \\ & & \ddots\end{array}\right),\end{equation}which for a single spin is precisely the state we have obtained above. The von Neumann entropy is invariant under unitary transformations and therefore does not change under Hamiltonian dynamics. Additionally, it is subadditive with respect to its subsystems, i.e.,
\begin{equation}S(\rho_{AB}) \leq S(\rho_A) + S(\rho_B),\end{equation}where $\rho_A$ and $\rho_B$ are the density matrices of the individual subsystems $A$ and $B$ \cite{Araki1970}. Consequently, by looking only at parts of the full system, we can only obtain partial information, as seen in the example above. Remarkably, for pure states of the full system, the entropies $S(\rho_A)$ and $S(\rho_B)$ are identical. This can be seen by looking at the Schmidt decomposition \cite{Nielsen2000} of the wave function of the full system,
\begin{equation}|\psi\rangle = \sum\limits_i c_i|\phi_i\rangle|\chi_i\rangle,\end{equation}which results in the reduced matrices
\begin{equation}\rho_A = \sum\limits_i |c_i|^2 |\phi_i\rangle\langle\phi_i|,\;\;\;\;\rho_B = \sum\limits_i |c_i|^2 |\chi_i\rangle\langle\chi_i|.\end{equation}As the coefficients $|c_i|^2$ are identical for both subsystems, the reduced density matrices have the same nonzero eigenvalues and therefore the same entropy. Therefore, we can define an ``entanglement entropy'' for any pure state $|\psi\rangle$ simply as the entropy of the reduced density matrix $S(\rho_A)$. In general, the entanglement entropy will depend on the choice of the subsystems, however, we are mostly interested in the partitioning that maximizes the entanglement entropy. For translationally invariant systems, this is achieved simply by cutting the system in half.

Coming back to the question on the usefulness of DMRG, we take a look at the entanglement entropy of matrix product states. As the states are encoded in terms of a $D\times D$ matrix, the entanglement entropy of matrix product states satisfies the relation
\begin{equation}S(\rho_A) \leq 2\log D,\end{equation}
which does not change with the size of the subsystem. So how does this figure compare to typical ground states of many-body systems? For gapped one-dimensional systems with local interactions, it can be shown that the entanglement entropy of the ground state is constant, which can be understood as a consequence of a finite speed of information transfer \cite{Eisert2010}. Consequently, for this large class of systems, DMRG can be expected to produce accurate results. Furthermore, for critical (gapless) models or models with strong disorder, one typically finds a logarithmic divergence of the entanglement entropy with system size, which still allows for an efficient simulation with DMRG \cite{Eisert2010}. On the other hand, in higher-dimensional systems, the entanglement entropy typically satisfies a relation of the form
\begin{equation}S(\rho_A) \lesssim \mathcal{A}(A),\end{equation}
where $\mathcal{A}(A)$ is the surface area of the subsystem \cite{Eisert2010}. Thus, the constant scaling of the entanglement entropy in one-dimensional systems can be seen as the surface area being constant in this case. The consequences of this ``area law'' of the entanglement entropy for DMRG are disastrous: we cannot expect DMRG to be a useful method for anything beyond one-dimensional systems. While there have been various approaches to extend DMRG-style methods to higher dimensions, their success has been fairly limited.

\bibliographystyle{apsrev} 
\bibliography{/home/itp/weimer/hendrik}


\title{Chapter 5: Digital quantum simulation}

\maketitle

\section{Quantum logic gates}

While the possibilities for quantum simulation based on ultracold atoms outlined previously have great potential, they are also somewhat constrained by the limits imposed by the contact interaction. In principle, it is possible to also include different interaction mechanisms such as the long-ranged dipole-dipole interaction \cite{Lahaye2009}, but as the ultimate goal, we would like to have a quantum simulator capable of simulating ''any'' quantum system. For practical purposes, we restrict the definition of such a universal quantum simulator to the simulation of Hamiltonians with short-ranged interactions \cite{Lloyd1996}, as all physical interactions reduce to purely local interaction at some point. Similar to numerical simulations on classical computers, it is also helpful to introduce abstraction layers so that we can use suitable approximations for the inner workings of the quantum simulator. If we decide to ignore the actual physical implementation for now, the lowest level we can consider is given by quantum logic gates, which have originally been discussed in the context of quantum computing \cite{Nielsen2000}.

Quantum logic gates are represented by unitary matrices that transform the quantum state before the operation into another after the application of the quantum gate. As such, they can be seen as the time-evolution operator acting for discrete timesteps,
\begin{equation}|\psi(\tau_{n+1})\rangle = U(\tau_n,\tau_{n+1})|\psi(\tau_n)\rangle.\end{equation}The basis set for the quantum state $|\psi\rangle$ is given by a product basis of two-level systems (quantum bits or ``qubits''),
\begin{equation}|\psi\rangle = \sum_{i \in \{0,1\}, j \in \{0,1\},\ldots} c_{ij\ldots} |ij\ldots\rangle.\end{equation}We can also think of the operation $U$ to be constructed out of several smaller building blocks,
\begin{equation}U(\tau_n,\tau_{n+1}) = \prod_i^N U(\tau_{n+(i-1)/N},\tau_{n+i/N}).\end{equation}For simplicity, we want to restrict ourselves to a universal set of quantum gates that can be used to construct any other gate from it. This can be realized by a set of three quantum gates, including the $z$ rotation gate,
\begin{equation}R_z(\phi) = \exp(i\phi\sigma_z) = \left(\begin{array}{cc}e^{i\phi} & \\ & e^{-i\phi}\end{array}\right),\end{equation}where $\phi$ is an arbitrary rotation angle. To construct any other single qubit quantum gate, we need a second gate that does not commute with $R_z$. The most convenient choice is the Hadamard gate given by
\begin{equation}U_H = \frac{1}{\sqrt{2}}\left(\begin{array}{cc}1 & 1\\1 & -1\end{array}\right).\end{equation}The Hadamard gate can be used to transform $\sigma_z$ into $\sigma_x$ and vice versa, i.e.,
\begin{eqnarray}R_x(\phi) &=& \exp(i\phi\sigma_x) = \left(\begin{array}{cc} \cos \phi & i\sin\phi  \\ i\sin\phi & \cos\phi\end{array}\right) = U_H R_z(\phi) U_H\\R_z(\phi) &=& U_H R_x(\phi) U_H.\end{eqnarray}Note that while the Hadamard gate is Hermitian, $U_H^\dagger = U_H$, most quantum gates are not. Finally, rotations about the $y$ axis can be constructed as
\begin{equation}R_y(\phi) = \exp(i\phi\sigma_y) = R_z(-\pi/4)R_x(\phi)R_z(\pi/4).\end{equation}Single qubit rotation do not allow us to generate entanglement between the qubits. Therefore, it is necessary to include a two-qubit quantum gate, which is most coveniently chosen as the controlled-not (CNOT) gate, acting on two qubits $A$ and $B$ as
\begin{equation}U_{CNOT} = |0\rangle\langle 0|_A \otimes 1_B + |1\rangle\langle 1|_A  \otimes \sigma^x_B = \left(\begin{array}{cccc}1\\ & 1\\& & & 1\\ & & 1\end{array}\right).\end{equation}Its function can be understood as follows: If the ``control'' qubit $A$ is in the 0 state, nothing happens. However, if $A$ is in 1, the ``target'' qubit $B$ gets flipped by the $\sigma_x$ operation. As an example, let us study the creation of entanglement between two qubits by a gate sequence consisting of a single Hadamard gate, followed by a CNOT operation. The qubits are initialized in the product state $|00\rangle$. Then, we have
\begin{equation}|\psi\rangle = U_{CNOT} U_H^A |00\rangle = \frac{1}{\sqrt{2}} (|00\rangle + |11\rangle),\end{equation}which is a maximally entangled state as its reduced density matrix is the maximally mixed state. The set of $R_z(\phi)$, $U_H$, and $U_{CNOT}$ forms a universal set for all $N$-qubit quantum gates \cite{Barenco1995}.

\begin{figure}[ht]
  \begin{tabular}{cc}
    \includegraphics{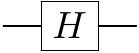} & \includegraphics{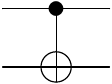}\\
    (a) & (b)
    \end{tabular}

\caption{Circuit representation of the Hadamard gate (a) and the CNOT gate (b).}

\end{figure}
It is often convenient to use a pictorial representation for quantum gate networks, also known as quantum circuits. Each qubit is represented by a straight line, with the horizontal axis denoting time, and each single qubit gate is shown a rectangular box acting on the particular qubit. Two-qubit gates are represented in a similar way, with the control qubit being indicated by a small circle, see Fig. 1. In this notation, much more complex quantum circuits can be represented and analyzed, for instance the decomposition of the Toffoli gate,
\begin{equation}U_T = \left(\begin{array}{cccccccc} 1 \\ & 1\\ & & 1\\ & & & 1\\ & & & & & 1\\ & & & & & & & 1\\ & & & & & & 1\end{array}\right),\end{equation}which is a 3-qubit variant of the CNOT gate and is universal for all classical computations. It can be constructed from elementary gates, see Fig. 2, where $T = R_z(\pi/8)$ \cite{Nielsen2000}.

\begin{figure}[ht]
  \includegraphics{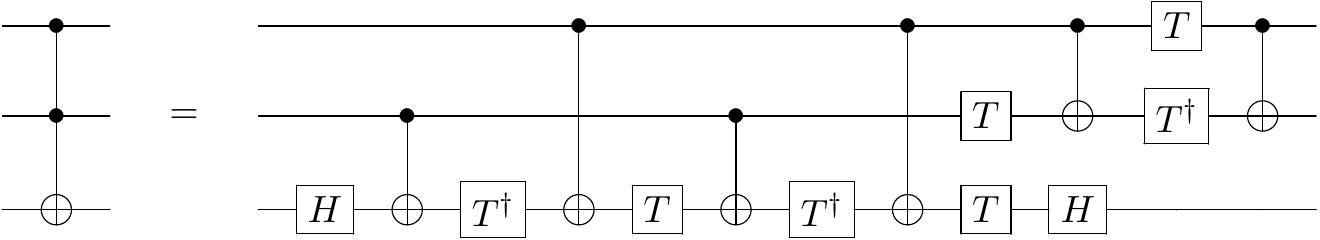}

\caption{Decomposition of the Toffoli gate into elementary quantum gates.}

\end{figure}

\section{Digital simulation procedure}

Suppose we want to simulate a four-body spin interaction of the form
\begin{equation}H=E_0 \sigma_x^{(1)}\sigma_x^{(2)}\sigma_x^{(3)}\sigma_x^{(4)}.\end{equation}This four-body spin operator has two eigenvalues, $\pm 1$, which are eightfold degenerate. The key idea is to use an additional auxiliary particle and encode the eigenvalue into its spin state. If the auxiliary control spin is initially in $|0\rangle$, this can be done using the gate sequence
\begin{equation}G = U_H^{(c)}\left(\prod_{i=1}^4 U_{CNOT}^{(c,i)}\right) U_H^{(c)}.\end{equation}To understand this in more detail, let us look at the behavior of the gate sequence on the spin state
$|\pm 1,\lambda\rangle$, where $\lambda$ labels the state within the degenerate manifold. The first Hadamard gate will yield
\begin{equation}U_H^{(c)}|0\rangle_c |\pm 1,\lambda\rangle = \frac{1}{\sqrt{2}}(|0\rangle_c+|1\rangle_c)|\pm 1,\lambda\rangle.\end{equation}Applying the sequence of CNOT gates will multiply the eigenvalue of the spin interaction, conditional on the control spin being in $|1\rangle$,
\begin{equation}\prod_i U_{CNOT}^{c,i} \frac{1}{\sqrt{2}}(|0\rangle_c+|1\rangle_c)|\pm 1,\lambda\rangle = \frac{1}{\sqrt{2}}(|0\rangle_c\pm|1\rangle_c)|\pm 1,\lambda\rangle.\end{equation}Finally, the second Hadamard gate will give us
\begin{eqnarray}U_H^{(c)}\frac{1}{\sqrt{2}}(|0\rangle_c+|1\rangle_c)|+1,\lambda\rangle &=& |0\rangle_c|+1,\lambda\rangle\nonumber\\U_H^{(c)}\frac{1}{\sqrt{2}}(|0\rangle_c-|1\rangle_c)|-1,\lambda\rangle = |1\rangle_c|-1,\lambda\rangle.\end{eqnarray}
Consequently, we have mapped the eigenvalue of the four-body interaction operator onto the state of a single auxiliary spin.

The full quantum simulation of the dynamics $U=\exp(-iHt)$ can then be realized by applying a $z$ rotation to the control spin and reverse the mapping $G$,
\begin{equation}U = \exp(-iE_0\sigma_x^{(1)}\sigma_x^{(2)}\sigma_x^{(3)}\sigma_x^{(4)}t) = G R_z(-\phi) G.\end{equation}The phase of the $z$ rotation is related to the timescale of the simulation according to $\phi = E_0 t$.

For a many-body system, the full dynamics can be simulated if the gate sequences are applied in parallel (if they act on independent spins) or sequentially (if they act on the same spins). However, in case of non-commuting operators, one has to ensure that this sequential operations does not introduce errors. This is true if the timestep $\tau$ of the simulation procedure is sufficiently small, as can be seen from the Suzuki-Trotter expansion
\begin{equation}\exp[-i(H_A+H_B)\tau] = \exp(-iH_A\tau)\exp(-iH_B\tau) + O(\tau^2).\end{equation}This completes the toolbox required for the realization of a universal quantum simulator.

\section{Implementation based on Rydberg atoms}

Let us now turn to a possible realization of such a universal quantum simulator based on ultracold Rydberg atoms \cite{Weimer2010}. The qubit states are formed by two hyperfine ground states. Single-qubit gates can be implemented using microwave driving described by the Hamiltonian
\begin{equation}H = \left(\begin{array}{cc}0 & e^{i\phi}\Omega\\e^{-i\phi}\Omega & \Delta\end{array}\right),\end{equation}where $\Delta$ is the detuning of the microwave frequency from the resonance between the hyperfine levels, $\Omega$ characterizes the strength of the microwave field, and $\phi$ describes the phase of the microwave field at $t=0$. This allows for the realization of arbitrary single-qubit gates.

For two-qubit gates, we need an interaction mechanism between the qubits. Ideally, we want the qubits to be separated by more than $500\,\mathrm{nm}$, so they can be addressed independently using optical laser fields. Hence, we can think of the atoms being localized in individual sites of an optical lattice, forming the $n=1$ Mott insulator. At such separations, however, the van der Waals interaction between ground state atoms is completely negligible. For dramatically increased interactions strengths, we will excite the atoms into a Rydberg state, which is an electronically excited state with a principal quantum number $n>10$. In these highly excited states, the atoms behave almost hydrogen-like, and their eigenenergies are given by
\begin{equation}E = -\frac{1}{2(n-\delta_l)^2},\end{equation}where $\delta_l$ is the quantum defect that accounts for deviations from the energy levels of hydrogen. For example, in rubidium in a $l=0$ state, the quantum defect has been measured to be $\delta_0 = 3.1311$ \cite{Mack2011}.

In Rydberg states, the excited electron is only very loosely bound and therefore very sensitive to external perturbations. This is also true if the perturber is another Rydberg atom and leads to a dramatic increase in the van der Waals interaction coefficients, scaling as $C_6 \sim n^{11}$. At the same time, the lifetime limited by spontaneous emission also increases as $\tau \sim n^3$, leaving enough time to perform a quantum gate before the Rydberg atom decays. The strong interaction between Rydberg atoms leads to a blockade mechanism: if one atom has been excited to a Rydberg state, its neighbors can no longer be excited at the same time as the interaction energy changes the resonance condition \cite{Jaksch2000}. To implement a CNOT gate, one first excites the $|0\rangle$ state of the control atom to the Rydberg state. Then, one tries to excite the target atom to the Rydberg state as well, which will only work if there are no interactions (i.e., the control atom is in $|1\rangle$), allowing for the conditional dynamics required for the implementation of the CNOT gate. The total gate requires a total of seven laser pulses and has been experimentally demonstrated in 2010 \cite{Isenhower2010}. Scaling up to the case of many qubits required for a full-fledged quantum simulator is currently underway.

\bibliographystyle{apsrev} 
\bibliography{/home/itp/weimer/hendrik}